\documentclass[prb.,twocolumn]{revtex4}
\usepackage{caption}
\usepackage{graphicx}
\usepackage{epsfig}
\usepackage{amsmath}
\usepackage{amsfonts}
\usepackage{amssymb}
\usepackage[version=3]{mhchem}
\providecommand{\U}[1]{\protect\rule{.1in}{.1in}}

\begin{document}

\title{\ Chiral current generation in QED by longitudinal photons }
\author{J. L. Acosta Avalo$^{*}$ and H. P\'{e}rez Rojas$^{**}$ ,}

\affiliation{$^{*}$\textit{Instituto Superior de Tecnolog\'{i}as y Ciencias Aplicadas  (INSTEC), Ave Salvador Allende, No. 1110, Vedado, La Habana, 10400 Cuba.}}

\affiliation{$^{**}$\textit{Instituto de Cibern\'{e}tica, Matem\'{a}tica y
F\'{\i}sica (ICIMAF), Calle
E esq 15, No. 309, Vedado, La Habana, 10400 Cuba. }}

\begin{abstract}
We report the generation of a pseudovector  electric current having imbalanced chirality in an electron-positron strongly magnetized gas in QED. It propagates along the external applied magnetic field $\textbf{B}$ as a chiral magnetic effect in QED. It is triggered by a perturbative electric field parallel to $\textbf{B}$,  associated to a pseudovector longitudinal mode propagating along $\textbf{B}$. An electromagnetic chemical potential was introduced, but our results remain valid even for vanishing chemical potential. A nonzero fermion mass was assumed, which is usually considered vanishing in the literature. In the quantum field theory formalism at finite temperature and density, an anomaly relation for the axial current was found for a medium of massive fermions. It bears some analogy to the Adler-Bell-Jackiw anomaly. From the expression for the chiral current in terms of the photon self-energy tensor in a medium, it is obtained that electrons and positrons scattered by longitudinal photons (inside the light cone) contribute to the chiral current, as well as  the pair creation due to longitudinal photons (out of light cone). In the static limit, an electric pseudovector current is obtained in the lowest Landau level.

\bigskip

\end{abstract}

\maketitle

\section{Introduction}
\label{sec1}

Nowadays the influence of the magnetic fields in relativistic quantum systems,
like electron-positron plasma and quark-gluon plasma (QGP) \cite{Lee}, is an important subject for its diverse applications. In \cite{Kha1} it was shown that a magnetic field in the presence of imbalanced chirality induces a current along the magnetic field. This is called Chiral Magnetic Effect in a QGP \cite{Kha2,Kha3}, which is caused by the topological gluon field configurations that couple to quarks via the QCD axial anomaly \cite{S1,Ioffe1,Hooft,Hooft1,Belavin,Manton}. The chiral magnetic effect in a QGP has important applications in the experiments of heavy ion collisions. An active experimental program exists to investigate the properties of this hot phase of matter (QGP) by using the Relativistic Heavy Ion Collider (RHIC) and the Large Hadron Collider (LHC).

In several papers about of the chiral magnetic effect the chirality is introduced through a nonvanishing chiral chemical potential factor \cite{Kha2,Alekseev}. In our paper, as done in \cite{Vilenkin}, we introduced an electromagnetic chemical potential $\mu$, but our results remain valid for $\mu=0$ (as it happens in the chiral magnetic effect in QCD). On the other hand, to obtain a full understanding of the chiral magnetic effect, it is desirable to include the dynamics leading to a net chirality. In our paper the pseudovector longitudinal mode propagating along $\textbf{B}$ is associated to the chiral charge density in a massive charged medium in the context of QED.

Usually, in current literature for the axial anomaly, it is assumed zero fermion mass \cite{Alekseev,JF}. This has the advantage that particles have definite helicity, but as pointed out by \cite{Vilenkin} the approximation is not realistic. For instance, in the presence of magnetic fields, it has unavoidable difficulties, since dealing with charged massless particles implies the arising of divergent magnetic moments.

As different from \cite{Vilenkin}, we start by establishing a difference between the chiral symmetry breaking due to nonzero mass (which we may name as scalar or dynamical), compatible with thermodynamical equilibrium (since in it at each instant an equal number of left and right particles are expected to be on the average) and the pseudoscalar chiral symmetry breaking, arising from the nonvanishing term $\mathfrak{G} =\frac{1}{4}\textbf{E} \cdot \textbf{B}$, which leads to  non-equilibrium processes of electric current and transport of charge. In what follows we want to show that a chiral magnetic effect arises in QED due to an anomaly relation for the axial current in a medium (not in vacuum: it is a purely quantum magnetic effect in a medium) of massive and magnetized charged fermions. The axial character will refer to the average density in momentum space of those particles and antiparticles having a common helicity. If  a perturbative electric field $\textbf{E}$, associated to a longitudinal pure electric mode (pseudovector mode), is applied to the electron-positron system in thermodynamic equilibrium in the presence of a very strong external magnetic field  $\textbf{B}$, such that $\textbf{E}\parallel\textbf{B}$, it produces an axial current leading to the breaking of the previously existing statistical chiral balance of  the densities of charged particles.

The existence of huge magnetic fields in {\it compact objects} is an established fact (typical measured surface magnetic fields are about $10^{12}$ G, although in the case of the {\it magnetar} subclass, it is believed that they can be as high as $10^{15}$ G \cite{Woods}). Therefore these objects are the best place to find direct experimental evidence of the chiral magnetic effect in magnetized charged plasmas. The electric current along $\textbf{B}$, associated to a chirality imbalance induced by longitudinal photons, could has astrophysical implications \cite{Ferrer,Miransky,JCAP}.

We remind that in a medium, as different from vacuum, there is a nonvanishing four-velocity vector $u_{\mu} \neq 0$. We must recall that also in absence of external fields in the charged medium there are three electromagnetic modes, two transverse and one longitudinal \cite{Fradkin2}. The two transverse modes correspond to photon spin projections  $\pm 1$ along its momentum $\bf{k}$, their dispersion equations being not on the light cone. The  third mode is a pure electrical longitudinal one (zero spin) and its dispersion equation  in the infrared limit has a solution for $\omega=0, \textbf{k}\neq 0$  which accounts for the Debye mass screening \cite{Fradkin2}, having close formal analogy to the Yukawa force.

In a quantum relativistic electron-positron plasma under the action of an external field $\textbf{B}$ generated by a four-potential $A_{\mu}^{ext}$, the photon
self-energy tensor structure determine the modes of propagation in both the $C$-invariant and non-invariant cases
\cite{Hugo2,Hugo3,Shabad,Shabad1}, as well as the electric currents. In our magnetized pasma is assumed a ionic positive background for the case $\mu \neq 0$ to guarantee total charge neutrality, although other effects can be neglected. The magnetic field breaks the spatial symmetry, leaving  invariances for rotations around $\bf{B}$, and for space translations along it. This determines different eigenmodes
according to the direction of propagation and the linear in the electric field expression of the currents. The dispersion equations for
photons propagating in the medium can be solved in any direction. For instance, the case of propagation parallel to $\textbf{B}$, in which we are interested \cite{Hugo3}, for transverse modes it was found the relativistic Hall conductivity as well as the Faraday Effect \cite{Lidice,Hugo6}. The last arise due to the breaking of the electromagnetic wave chiral symmetry induced by the $C$-non-invariance of the transverse modes. In our case, chiral symmetry is induced by
the breaking of electric field symmetry along $\textbf{B}$.

The conductivity associated to the pseudovector mode is proportional to the corresponding eigenvalue of the photon self-energy tensor in the medium. We  discuss the structure of this current in terms of  the scattering and pair creation of electrons and positrons resulting from the decay of the longitudinal photons. We show that a current is produced along $\textbf{B}$ in the presence of imbalanced chirality (induced by longitudinal photons), separating charges of opposite sign but the same helicity. We consider this effect as a chiral magnetic effect in the context of QED for its analogy with the QCD chiral magnetic case. These results could be extended also to a quark-antiquark system in a strong magnetic field by taking into account their electromagnetic interaction.

The present results are based on the general properties of the propagation of electromagnetic waves, and the electromagnetic current vector, in the presence of the field $\textbf{B}$, and we use the quantum field theory formalism at finite temperature \cite{Fradkin2} $T \neq 0$ ($T$ is given in energy units) and density $\mu \neq 0$ (the medium is $C$-non invariant)\cite{Hugo2,Hugo3}.

In the external constant field $\textbf{B}$ (which we assume as parallel to the $x_3$ axis), electrons and positrons move in bound states characterized by energy levels  given by (we will use $\hbar=c=1$ units, except for specific calculations):

\begin{equation}\label{espectro/energia}
   \varepsilon_{n_{l},p_{3}}=\sqrt{p^{2}_{3}+m^{2}+|e|B(2n_{l}+1-sgn(e) s_{3})},
\end{equation}

\noindent where $p_3$ is the momentum along $B$, $m$ is the electron mass, $s_{3}=\pm 1$ are spin eigenvalues along $x_{3}$ and $n_{l}=0,1,...$ are the Landau quantum numbers. These are two-fold spin degenerate, except the ground state $\varepsilon_{0}$ in which $n_{l}=0$, and for electrons it is $s_{3}=-1$ and for positrons $s_{3}= 1$.  Quantum states
are also degenerate with regard to the orbit's center coordinates \cite{John}. If we consider strong magnetic fields, such that $2eB \gg \mu^2,T^{2}$, the contribution of the lowest Landau level (LLL) becomes dominant, whereas the contribution from higher Landau levels is negligibly small. Thus, for a system in equilibrium at temperature $T$ and chemical potential $\mu$, the net density of charged particles in the  ground state $N_0=-\partial \Omega/\partial\mu$  can be written from the linear in $B$ thermodynamic potential $\Omega$ \cite{prl2000}, as:

\begin{equation}\label{densidad-particulas/basico}
  N_0=\frac{eB}{2\pi^{2}}\int_{-\infty}^{0}dp_3 (n^{e}_{R}-n^{p}_{L}) + \int_{0}^{\infty}dp_3(n^{e}_{L}-n^{p}_{R}),
\end{equation}

\noindent and its magnetization ${\cal M}_0=-\partial \Omega/\partial B (\equiv -\Omega/B)$ by:

\begin{equation}\label{magnetizacion/basico}
  {\cal M}_0=\frac{e}{4\pi^{2}}\int_{-\infty}^{0}\frac{p_3^2 dp_3}{\varepsilon_0} (n^{e}_{R}+n^{p}_{L}) + \int_{0}^{\infty}\frac{p_3^2 dp_3}{\varepsilon_0}(n^{e}_{L}+ n^{p}_{R}),
\end{equation}

\noindent where $n^{e,p}=[1 + e^{(\varepsilon_{0} \mp \mu)/T}]^{-1}$ are the electron and positron densities in momentum space, and in Eq. (\ref{densidad-particulas/basico}), Eq. (\ref{magnetizacion/basico}) their left- and right-handed helicities are labeled as $L,R$ respectively. According to these formulae the magnetic moments are aligned along $\textbf{B}$, showing a paramagnetic behavior. Notice that, by changing $p_3 \to -p_3$, both terms in each sum  are  equal, which means $n^{e,p}_{R,L}=n^{e,p}_{L,R}$ in the state of equilibrium. As the magnetization ${\cal M}_0>0$, it implies that left electrons and right positrons have positive momentum and  move parallel to $\textbf{B}$ whereas right electrons and left positrons having negative momentum, move antiparallel to it. \textit{If an
external perturbative electric field $\textbf{E}$ is applied along $\textbf{B}$, it supplies external momentum to the system, and thermodynamic equilibrium as well as chiral balance are broken}. This leads to a chiral current, that is, a chiral magnetic effect in QED. The charges will move according to the relative directions of $\textbf{E}$ and $\textbf{B}$. In the case they are parallel, electrons  decrease their negative momentum and  positrons increase their positive momentum (if $T=0$, positrons have a negligible contribution, and for electrons the Fermi "sphere" is shifted up), leading to $n^{e,p}_{R}> n^{e,p}_{L}$, producing a chiral R-current, for the antiparallel case we have the inverse effect, leading to an L-current, where the sign of the pseudoscalar $\mathfrak{G} = \frac{1}{4}\textbf{E} \cdot \textbf{B}$ plays a fundamental role. This gives a phenomenological basis to our chiral magnetic effect. A quantitative approach is given below.

\section{Chiral current generation in QED}
\label{sec2}

The electromagnetic current may be interpreted as generated from the decay of longitudinal photons in scattering of electrons and positrons and pair creation. The Schwinger-Dyson equation for the photon in Fourier space, with $\nu=1,2,3,4$, is \cite{Fradkin2}:

\begin{equation}\label{ecuacion/SD}
  [k^{2} g_{\mu\nu}-\Pi_{\mu\nu}(k|A_{\mu}^{ext})]a^{\nu}(k)=0,
\end{equation}

\noindent where $g_{\mu\nu}$ is the Minkowski metric, in the form $g_{\mu\nu}=(1,1,1,-1)$ (it corresponds to the analytic continuation $k_{4}\rightarrow i\omega$ from Euclidean metric), and $k^{2}=k_{3}^{2}+k_{\bot}^{2}-\omega^{2}$. Here $k_{3}$ and $k_{\bot}$  are respectively the components of the photon four-momentum in directions parallel and perpendicular to $\textbf{B}$, and $\omega$ its energy. The total external electromagnetic field is  $A^{ext}_{\mu}+a_{\mu}$, where $a_{\mu}$ is a small perturbative radiation field (its electric field  $E\ll B$). The quantum corrections are given by the photon self-energy tensor $\Pi_{\mu\nu}(k|A_{\mu}^{ext})$, which was calculated in magnetized vacuum and in the one-loop approximation in  \cite{Shabad1}, and in a magnetized medium in \cite{Hugo2,Hugo3,Shabad}.

According to \cite{Hugo2,Shabad1}, the diagonalization of the photon self-energy tensor in vacuum and in a medium  leads to the
equation:

\begin{equation}\label{Op/autovectores}
 \Pi_{\mu\nu} b^{\nu(i)}=\eta_{i}b_{\mu}^{(i)},
\end{equation}

\noindent  having
three non-vanishing eigenvalues $\eta_{i}$ and three eigenvectors
$b_{\mu}^{(i)}$ for $i = 1, 2, 3,$ corresponding to three photon
propagation modes, whose electric and magnetic fields were obtained in
\cite{Shabad1}. The photon
four-vector $k_{\nu}$ has zero eigenvalue. For each mode it is obtained a dispersion law
$k^{2}=\eta_{i}(k_{3},k_{\bot},\omega,B)$ \cite{Hugo2,Hugo3,Shabad,Shabad1}.

In a charged $e^{\pm}$ medium, for propagation along the field $\textbf{B}$,  in addition to the two transverse modes (see Eq. (\ref{eigenmodes}) in \ref{sec8}), there is a longitudinally polarized mode along $\textbf{B}$ given by the pseudovector:

\begin{equation}\label{pseudo-vector}
  b_{\mu}^{(2)}(k)=a c_{\mu}^{(2)},
\end{equation}

\noindent where $c_{\mu}^{(2)}=R_{2}(F^{*}k)_{\mu}$ is a normalized pseudovector  (the normalization parameter is $R_{2}=1/Bz_{1}^{1/2}$), and $F^{*}_{\mu\nu}$ is the dual of the electromagnetic field tensor $F_{\mu\nu}$. The parameter $a$ (which has dimension of vector potential) is determined by the applied perturbative electric field. Its electric polarization vector being in the direction along $\textbf{B}$ \cite{Shabad1}

\begin{equation}\label{campo-electrico}
  \textbf{E}_{B}=E^{(2)}\textbf{e}_{B}=a(k_{3}^{2}-\omega^{2})^{\frac{1}{2}}\textbf{e}_{B},
\end{equation}

\noindent where $\textbf{e}_{B}=\textbf{B}/B$ is a unit pseudovector. As pointed out earlier,
the longitudinal mode is not on                                                                                                                                                                                                                                                                                                                                                                                                                                                                                                                                                                                                                                                                                                                                                                                                                                                                                                                                                                                                                                                                                                                                                                                                                                                                                                                                                                                                                                                                                                                                                                                                                                                                                                                                                                                                                                                                                                                                                                                                                                                                                                                                                                                                                                     the light cone, that is $k^{2}_{3}-\omega^{2}\neq 0$ \cite{Shabad1}. From now on we will call  $z_{1}=k_{3}^{2}-\omega^{2}$.

The electromagnetic current as a function of $A^{ext}_{\mu}+
a_{\mu}$ depends on the two relativistic invariants:

 \begin{equation}\label{invariante1}
     \mathfrak{F}= \frac{1}{4}F^{\mu\nu}F_{\mu\nu} =\frac{1}{2}(B^{2}-E^{2})\simeq\frac{1}{2}B^{2}
 \end{equation}
\begin{equation}\label{invariante2}
  \mathfrak{G} = \frac{1}{4}F^{*\mu\nu}F_{\mu\nu}=\textbf{B}\cdot\textbf{E}.
\end{equation}

Notice that for the case of propagation along $\textbf{B}$, the pseudoscalar $\mathfrak{G}\neq0$ only for the longitudinal mode $b_{\mu}^{(2)}$, independently of the $C$-symmetric of the system. An expansion of the electromagnetic current density in functional series of $a_\nu$ gives:

\begin{align}\label{desarr-corriente}
  j_{\mu}(A^{ext}_{\mu}+
a_{\mu})=j_{\mu}(A^{ext}_{\mu}) + \frac{\delta j_{\mu}}{\delta A_{\nu}^{ext}}a_{\nu}+... \hspace{2mm} ,
\end{align}

\noindent its linear term in $a_{\nu}$ is \cite{Lidice,Hugo6}:

\begin{align}\label{Op}
  j_{i}=\Pi_{i\nu}a^{\nu}=Y_{ij}E_{j},
\end{align}

\noindent where $E_{j}=i(\omega a_{j}-k_{j}a_{0}) $ is the electric field, with $a_{4}=ia_{0}$ and $k_{4}=i\omega $, also $j_{\mu}(A^{ext}_{\mu})= N_{0}\delta_{\mu 4}$. The term
$Y_{ij}=\Pi_{ij}/i\omega$ is the complex conductivity tensor or
admittivity. The third term in Eq. (\ref{Op}) comes from the second
one by using the four-dimensional transversality of $\Pi_{\mu\nu}$
due to gauge invariance, $\Pi_{\mu\nu}k^{\nu}=0$ \cite{Hugo2,Hugo3,Shabad,Shabad1}. In Eq. (\ref{desarr-corriente}) $a_{\mu}$ is in general a linear function of the eigenmodes $b_{\mu}^{(i)}$. Below we particularize to the case in which the eigenvector
$a_{\mu}=b_{\mu}^{(2)}$,  for which the electric field vector is parallel to $\bf{B}$ (notice that only terms contaning odd number of $b_{\mu}^{(2)}$ legs in (\ref{desarr-corriente}) lead to pseudovector terms).

Charged fermions interacting with the longitudinal mode, exchange energy by the transfer of momentum $k_3$, while the Landau quantum numbers remain unchanged \cite{Hugo6}. Then we may consider the fermion interaction with the longitudinal mode as a problem in $(1+1)$ dimensions, which is strictly valid if we consider only the LLL. We would like to point out that the two-dimensional Dirac matrices obey the identity \cite{Peskin}:

\begin{align}\label{identidad-axial}
   \gamma^{\mu}\gamma^{5}=-\epsilon^{\mu\nu}\gamma_{\nu}.
\end{align}

This implies that the axial $j_{A}^{\mu}$ and vector $j^{\mu}$ currents exchange their $(0,3)$ components according to the same relation. Thus, in the $(1+1)$ case, we can study the properties of the axial vector current by using results already derived for the vector current.

Now we must observe that in the linear-in-$E_{i}$ approximation of $j_{i}$ Eq. (\ref{Op}), and from the eigenvalue equation Eq. (\ref{Op/autovectores}) one gets also:

\begin{equation}\label{chir}
  j_{i}=\Pi_{i\nu}a^{\nu}=s b_{i}^{(2)},
\end{equation}

\noindent where we can write the scalar $s = c^{(2)\mu}\Pi^{\nu}_{\mu}c^{(2)}_{\nu}$, which is the eigenvalue of the photon self-energy tensor corresponding to the longitudinal mode \cite{Hugo2,Hugo3,Shabad}. The remarkable fact is that as $b_{\nu}^{(2)}$ is a pseudovector, \textit{for propagation along $\textbf{B}$ the current $j_{\nu}$ is also a pseudovector, which is a necessary condition for the breaking of chiral symmetry}.

It is easy to find a gauge transformation (in which it is obtained $b_{3}^{(2)}=(k_{4}/z_{1}) E_3$) leading to $j_3=s(k_{4}/z_{1}) E_3$ (from Eq. (\ref{chir})), where $E_{3}= E^{(2)}(\textbf{e}_{B}\cdot\textbf{e}_{3})$ . This equation is equivalent to
\begin{equation}\label{jota3}
j_{3}=\frac{\Pi_{33}}{k_{4}}E_{3},
\end{equation}

\noindent which is deduced from relation $\Pi_{33}=s(k_{4}^{2}/z_{1})$, it is obtained from the expression $s = c^{(2)\mu}\Pi^{\nu}_{\mu}c^{(2)}_{\nu}$, and from the two-dimensional tranversality $\Pi_{\mu\nu}k^{\nu}=0$, where $\mu, \nu=3,4$.

We are interested only in the real part of $j_3$, as we will restrict ourselves to the imaginary part of the photon self-energy tensor. From Eq. (\ref{identidad-axial}), Eq. (\ref{chir}) and by using the two-dimensional transversality condition of $\Pi_{\mu\nu}$, it is obtained:

\begin{align}\label{anomalia axial}
   k_{\mu}j^{\mu}_A=\frac{z_{1}}{k_{4}}j_{3}\neq0 ,
\end{align}

\noindent while $k_{\mu}j^{\mu}=0$, which expresses the conservation law for the vector current. Eq. (\ref{anomalia axial}) expresses the non-conservation of the two-dimensional axial current, whereas Eq. (\ref{jota3}) puts in evidence the role of the electric field, characterizing the longitudinal pseudovector mode, in the breaking of the chiral symmetry in both the $C$-symmetric and non-symmetric cases, which produces an electric current along $\textbf{B}$. This proves that a chiral magnetic effect is produced in the frame of QED. From now on we will restrict only to the real frequency and momentum ($k_{3}^{2}>z_{1}$).

\subsection{Chiral conductivity in a charged magnetized medium}
\label{sec3}

We are interested in the real conductivity which can be expressed explicitly in terms of the imaginary part of the photon self-energy tensor as \cite{Hugo2,Hugo3,Shabad,Shabad1}:

\begin{equation}\label{conduct-Real}
  \sigma_{ij}=\frac{Im[\Pi_{ij}]}{\omega}.
\end{equation}

The contribution to the current density  $j_{i}$ in Eq. (\ref{Op}) due to
conductivity can then be written in the general form as:

\begin{equation}\label{corriente/ohm}
j_{i}=\sigma_{ij}^{0}E_{j}+(E\times S)_{i},
\end{equation}

\noindent where $\sigma_{ij}^{0}=Im[\Pi_{ij}^{S}]/\omega$ and
$S_{i}=\frac{1}{2}\epsilon^{ijk}\sigma_{jk}^{H}$ is a pseudovector
associated with $\sigma_{jk}^{H}=Im[\Pi_{ij}^{A}]/\omega$,
$\epsilon^{ijk}$ is the third rank antisymmetric unit tensor.
$\Pi_{ij}^{S},\Pi_{ij}^{A}$ are the symmetric and antisymmetric
parts of the  photon self-energy tensor. The
first term corresponds to the
Ohm current and the second is the Hall current  \cite{Lidice}. The Hall and Faraday effects occur for the $C$-non-symmetric case; as different from the chiral magnetic effect, which occurs in both the $C$-symmetric and non-symmetric cases. Here we will only work
with the Ohm current associated to the  longitudinally  polarized
mode (which was not discussed in the earlier papers \cite{Hugo3}).

Let us consider the specific case of current along $\textbf{B}$, which is chiral non-symmetric. Form Eq. (\ref{corriente/ohm}) the current density associated to
the longitudinal mode can be expressed in the form:

\begin{equation}\label{corrient/conduct}
  j_{3}=\sigma_{33}^{0}E_{3},
\end{equation}

\noindent where $\sigma_{33}^{0}$ is the chiral conductivity associated to the longitudinal mode (recall that we defined earlier $E_{3}= E^{(2)}(\textbf{e}_{B}\cdot\textbf{e}_{3})$), which now will be calculated in a charged $e^{\pm}$ medium. From Eq. (\ref{conduct-Real}) and taking into account the above mentioned relation between $\Pi_{33}$ and $s$, it can be expressed as:

\begin{equation}\label{conductividad-escalar-s}
  \sigma_{33}^{0}=\frac{Im[\Pi_{33}]}{\omega}=-\frac{\omega Im[s]}{z_{1}}.
\end{equation}

The scalar $s$ in the one-loop approximation is
\cite{
Hugo2,Hugo3,Shabad,Shabad1}:

\begin{multline}\label{escalar-s}
  s\hspace{-0.1cm}=\hspace{-0.1cm} -\frac{e^{3}B}{\pi^{2}}\hspace{-0.1cm}\sum_{n=0}^{\infty}\int_{-\infty}^{\infty}\frac{dp_{3}}
  {\varepsilon_{q}}[\alpha_{n}\varepsilon^{2}_{n,0}(2p_{3}k_{3}+z_{1})]\\
  \times\frac{[n^{p}(\varepsilon_{q})+n^{e}(\varepsilon_{q})-1]}{4z_{1}p_{3}(p_{3}
  +k_{3})+z^{2}_{1}-4 \omega^{2} \varepsilon^{2}_{n,0}},
\end{multline}

\noindent where  $\varepsilon_{n,0}=\sqrt{m^{2}+2eBn}$, with $n=n_{l}+1/2+s_{3}/2$, $\alpha_{n}=2-\delta_{n,0}$ and $q=(n,p_{3})$. Here the term $-1$ inside the square brackets accounts for the quantum vacuum limit ($\mu=T=0$). From the expression for the imaginary part of the scalar $s$  obtained in \cite{Hugo3} (see \ref{sec9}), we have

\begin{align}\label{Imag-s}
Im[s]=-\frac{e^{3}B}{2\pi}\sum_{n=0}^{\infty}\alpha_{n}\varepsilon_{n,0}^{2}S_{n}.
\end{align}

From Eq. (\ref{conductividad-escalar-s}) and Eq. (\ref{Imag-s}), the chiral conductivity at a finite
temperature $T$ and density characterized by a chemical potential
$\mu$, it is found to be:

\begin{equation}\label{conductividad}
  \sigma_{33}^{0}=\frac{e^{3}B\omega}{2\pi z_{1}}\sum_{n=0}^{\infty}\alpha_{n}\varepsilon_{n,0}^{2}S_{n},
\end{equation}

\noindent here

\begin{equation}\label{Expre-S}
  S_{n}=\frac{1}{\Lambda}(\theta_{1}\Delta N+\theta_{2}\Delta H),
\end{equation}

\noindent where $\theta_{1}=\theta(z_{1})$ and $\theta_{2}=\theta(-4\varepsilon_{n,0}^{2}-z_{1})$ are the Heaviside step functions, and

\begin{equation}\label{exitacion}
 \Delta N=[N(\varepsilon_{r})-N(\varepsilon_{r}+\omega)],
\end{equation}

\begin{equation}\label{creacion/pares}
 \Delta H=[H(-\varepsilon_{s})+H(\omega+\varepsilon_{s})-2],
\end{equation}

\noindent where the term $N=n^{e}(\varepsilon_{r})+n^{p}(\varepsilon_{r})$ while the term $H=n^{e}(\varepsilon_{s})+n^{p}(\omega-\varepsilon_{s})$. Here $\Lambda=\sqrt{z_{1}(z_{1}+4\varepsilon_{n,0}^{2})}$, and $\varepsilon_{s}=(\omega z_{1}+|k_{3}|\Lambda)/2z_{1}$,
$\varepsilon_{r}=(-\omega z_{1}+|k_{3}|\Lambda)/2z_{1}$ with $r,s=(n,\omega,k_{3})$ are the
fermion energies in a magnetic field in terms of the longitudinal mode energy $\omega$ and momentum $k_{3}$. The term $\Delta N$ accounts for the excitation of particles $[\varepsilon (p_3,n)\longrightarrow \varepsilon (p_3+k_3,n)]$ by increasing their momentum along $\bf{B}$ (only in the region $z_{1}>0$), while $\Delta H$ accounts for the pair creation (only in the region $z_{1}<-4\varepsilon^{2}_{n,0}$), both due to the interaction with the longitudinal mode, the Landau quantum numbers being unchanged \cite{Hugo6}. Pauli's principle demands that vacant states must exist both for the occurrence of excitation and pair creation processes, for a fixed $n$. We will not consider the vacuum contribution in Eq. (\ref{conductividad}), which is associated with the term $-2$ in Eq. (\ref{creacion/pares}). Notice that from Eq. (\ref{anomalia axial}), Eq. (\ref{corrient/conduct}) and Eq. (\ref{conductividad}) is obtained that the pair creation process induces an imbalanced chirality (for $z_{1}<-4\varepsilon^{2}_{n,0}$), which produces an electric current along $\textbf{B}$. In the region $z_{1}>0$, only the scattering of electrons and positrons contribute to the electric current.

\subsection{Chiral current in the static limit for the LLL}
\label{sec4}

If we consider in the Eq. (\ref{conductividad}) only the contribution of LLL, the Landau states for $n>0$ is negligible small (strong magnetic fields such that, $2eB\gg\mu^{2},T^{2}$ ), it is obtained

\begin{equation}\label{conduct-est-bas}
 \sigma_{33}^{0}=\frac{e^{3}B\omega}{2\pi z_{1}}m^{2}S_{0},
\end{equation}

\noindent with

\begin{equation}\label{Expre-S-basico}
S_{0}=\frac{\theta_{1}\Delta N_{R}+\theta_{2}\Delta H_{R}}{\sqrt{z_{1}(z_{1}+4m^{2})}},
\end{equation}

\noindent where $\theta_{2}=\theta(-4m^{2}-z_{1})$. Notice that the particles must have right-handed helicity (for electric field parallel to $\textbf{B}$), then $\Delta N_{R}>(1/2)N_{0}$. The last fact is a consequence of the paramagnetic magnetization of the system (the higher (degenerate) Landau states with $R$-helicity are dominant, the  Eq. (\ref{conductividad}) contain both diamagnetic and paramagnetic terms). The current is a transport phenomenon, since it carries charge, and in consequence, a non-equilibrium process. Thus, the action of the external electric field $E_3$ is to break the equilibrium, and its more interesting  consequence is to break the chiral symmetry of the system.

At the low frequency limit $\omega \rightarrow0$, we have that $k_{3}\gg\omega$, that is, conditions very close to the static electric field case $(\omega=0)$. It must be in correspondence with the lower energy limit. The Eq. (\ref{Op}) is valid in this limit for the excitation of particles, which is guaranteed by the gauge invariance of $a_{\mu}$.

Taking into account Eq. (\ref{chir}), and doing an appropriate gauge transformation (in which it is obtained the pseudovector potential $b_{3}^{(2)}=E_3/k_4$, with $E_3=a|k_3|(\textbf{e}_{B}\cdot\textbf{e}_{3})$, from Eq. (\ref{campo-electrico}) in the low frequency limit), one gets

\begin{equation}\label{gauge}
j_3 =\frac{Im [s]}{\omega} E_3.
\end{equation}

Thus, as a result of Eq. (\ref{Imag-s}), only for the LLL, and Eq. (\ref{gauge}), we obtain

\begin{multline}\label{corriente-limite-estatico}
  j_{3}\hspace{-0.1cm}=\hspace{-0.1cm}
  \frac{e^3}{8\pi}\frac{m^{2}}{|k_{3}|\lambda}\frac{1}{T}\\
  \times[\frac{1}{1+\cosh(\frac{\lambda-\mu}{T})}+\frac{1}{1+\cosh(\frac{\lambda+\mu}{T})}](\textbf{E}^{(2)}\cdot \textbf{B}),
\end{multline}

\noindent where $\textbf{E}^{(2)}=a|k_3|\textbf{e}_{3}$, and $\lambda=(\sqrt{k_{3}^{2}+4m^{2}})/2$. If we consider low temperatures (positron contribution is negligible), and $mc\gg p_{3}$, with $\mu\gtrsim mc^2$ (we return to units $c,\hbar$) from Eq. (\ref{corriente-limite-estatico}) and multiplying by $\textbf{e}_{3}=\textbf{k}_3/k_3$, it is obtained

\begin{equation}\label{corriente-limite-estatico-temperatura-bajas}
  \textbf{j}_{3}=\frac{\alpha e}{16\pi}\frac{mc^2}{|p_3|}\hspace{1mm}\frac{e^{-|\frac{P_{3}^{2}}{2m}-\mu_{0}|/T}}{T}
   \hspace{1mm}(\textbf{E}^{(2)}\cdot\textbf{B})\textbf{e}_{3},
\end{equation}

\noindent where $\alpha$ is the fine-structure constant, and $\mu_{0}=\mu-mc^2$ in the non-relativistic chemical potential. In the static limit $\hbar k_{3}=2p_{3}$ (see Eq. (\ref{momentum}) in \ref{sec9}). The chiral current (Eq. (\ref{corriente-limite-estatico-temperatura-bajas}), which could be tested at the laboratory) bears some similarity with the one obtained in  \cite{Vilenkin}, but as different from it, Eq. (\ref{corriente-limite-estatico-temperatura-bajas}) refers to a massive medium, also at non-vanishing temperature, and chemical potential. It is  a peaked curve around $\mu_{0}$, which for $T\rightarrow 0$ degenerates proportional to the delta function $\delta(\mu_{0}-p_{3}^{2}/2m)$. A more general expression than  Eq. (\ref{corriente-limite-estatico-temperatura-bajas}), depending also of the frequency $\omega$, can be obtained from Eq. (\ref{corrient/conduct}) and Eq. (\ref{conductividad}).

\subsection{ Axial anomaly in the presence of mass in a magnetized medium }
\label{sec5}

We will find now an expression for the four-divergence of the axial current in a medium of massive particles in the presence of a field $\textbf{B}$. Taking into account Eq. (\ref{anomalia axial}) and  Eq. (\ref{conductividad}), with $\textbf{E}^{(2)}=E^{(2)}\textbf{e}_{3}$, and $\beta=-ie^{2}/2$, it is obtained:

 \begin{align}\label{corriente-anomalia}
  k_{\mu}j_{A}^{\mu}=\beta[m\mathbb{A}(m)+\mathbb{C}(m)]\frac{e^{2}}{2\pi^{2}}(\textbf{E}^{(2)}\cdot \textbf{B}),
\end{align}

\noindent where

\begin{equation}\label{coeficientes-1}
 \mathbb{A}(m)=\frac{2\pi m}{e}\sum^{\infty}_{n=0}\alpha_{n}S_{n},
\end{equation}

\begin{equation}\label{coeficientes-2}
  \mathbb{C}(m)=8\pi B\sum^{\infty}_{n=1}nS_{n}.
\end{equation}

Notice that the contribution to the non-conservation of the axial current Eq. (\ref{corriente-anomalia}) term may be due to excitation of particles or to the pair creation. The Eq. (\ref{corriente-anomalia}) is exactly an anomaly relation in a medium of massive particles in the presence of $\textbf{B}$, that bears some analogy  to the Adler-Bell-Jackiw (ABJ) relation in vacuum  \cite{Adler1,Bell}. The remarkable difference of Eq. (\ref{corriente-anomalia}) with the ABJ relation is that the pseudoscalar factor $\textbf{E}^{(2)}\cdot \textbf{B}$ appears also multiplying the usual massive term, vanishing only in the  $m \to 0$ limit. In this respect in a charged medium the longitudinal axial photon plays here a similar role than the $\pi^{0}$-meson at the vertex of Adler-Bell-Jackiw triangle.

A closer correspondence with the Adler-Bell-Jackiw triangular anomaly $(\pi^{0}\rightarrow\gamma\gamma)$ would be obtained by taking higher order terms in the expansion  Eq. (\ref{desarr-corriente}) by keeping only an odd number of longitudinal mode legs in each term. According to our present model, the triangle may be understood as due to the decay or absorption of the longitudinal axial photon in one vertex, by processes involving two transverse photons in the other vertices. Among these processes, we have the possibility of longitudinal photon splitting in two transverse ones in a charged $e^{\pm}$ medium under the action of a very strong field $\textbf{B}$ \cite{Baier}.

\section{Conclusions}
\label{sec6}

We conclude, that as a consequence of general properties of the spatial symmetry of a charged medium,  broken by a magnetic field $\textbf{B}$, the propagation of electromagnetic waves as well as the electric current parallel to it, leads to chiral effects. For the electromagnetic waves along $\textbf{B}$, we have
the Faraday effect, and for electric currents along $\textbf{B}$, the
so-called chiral magnetic effect.

If an external perturbative electric field $\textbf{E}$ is applied along $\textbf{B}$ in a dense medium, it supplies external momentum to the system, and thermodynamic equilibrium as well as chiral balance are broken via an axial relation. Thus, a chiral magnetic effect exists in QED, manifested in the induction of an electric pseudovector  current along $\textbf{B}$, separating massive fermions pairs of the same helicity, in analogy to the QCD chiral magnetic case.

In the QCD case the imbalanced chirality is due to the change in the topological charge (by introducing a chiral chemical potential in a medium of massless fermions \cite{Kha2,Alekseev}) whereas the imbalanced chirality in our massive charged medium is induced by the perturbative electric field in QED, due to the scattering of electrons and positrons or to the pair creation. This is the main difference between our QED chiral magnetic effect and the QCD chiral magnetic case reported in the literature.

The chiral conductivity was calculated in the general case of finite temperature and density, and for massive fermions, which might be relevant for astrophysical applications. From the  expression for chiral current in terms of photon self-energy tensor in a  magnetized medium, it is obtained that electrons and positrons scattered by longitudinal photons (inside the light cone) contribute to the chiral current, as well as the pair creation due to longitudinal photons (out of light cone). We also obtained an expression for the chiral current along $\textbf{B}$ in the static limit, which depends on the mass and momentum of the particles, and on the temperature and non-relativistic chemical potential.

We found a new anomaly relation for the axial current in a medium of massive particles in the presence of an external field $\textbf{B}$, that bears some analogy to the Adler-Bell-Jackiw (ABJ) relation in vacuum. This means the possibility of longitudinal photon splitting in two
transverse ones in a charged magnetized medium.

Finally, notice that the chiral effect that we have discussed in QED in a medium can be extended to quark-antiquark electromagnetic interactions, and even be combined with QCD effects. A separate study of this topic will be addressed in a future work

\section*{Acknowledgments}
\label{sec7}
The authors thanks the Abdus Salam ICTP OEA Office for support under Net-35 and to A. Cabo, A. P\'{e}rez Mart\'{i}nez and E. Rodr\'{i}guez Querts for fruitful discussions. We thank also A.E. Shabad for comments at the early stages of the present research.

%\appendix

\section{Appendix}

\subsection{Electromagnetic modes for propagation along \textbf{B}}
\label{sec8}

From the structure of $\Pi_{\mu\nu}$ in a magnetized plasma,
in the case of non-vanishing temperature $T$ and chemical
potential $\mu$ \cite{Hugo2,Hugo3}, we can find the polarization properties of three electromagnetic eigenmodes propagating in the system \cite{Hugo2,Shabad1}. Under those conditions the polarization tensor may be expanded in
 terms of six independent transverse tensors \cite{Hugo3}

 \begin{equation}\label{diagonalize form of the polarization tensor}
   \Pi_{\mu\nu}=\overset{6}{\underset{n=1}{\sum}}\pi^{(i)}\Psi_{\mu\nu}^{(i)}.
 \end{equation}

 As is shown in
  \cite{Hugo2}, symmetric properties in quantum statistics, reduce the number of the basic tensors from an initial set of $9$ to a final set of $6$. The basic tensors are

\begin{eqnarray}\label{tens-basicos-S}
  \Psi_{\mu\nu}^{(1)} &=& k^{2}T_{\mu\nu},\quad\Psi_{\mu\nu}^{(2)}=(Fk)_{\mu}(Fk)_{\nu}\nonumber\\
  \Psi_{\mu\nu}^{(3)} &=& -k^{2}T_{\mu}^{\lambda} F_{\lambda\eta}^{2}T^{\eta}_{\nu},\quad\Psi_{\mu\nu}^{(4)}=R_{\mu}R_{\nu},
\end{eqnarray}

\noindent with $T_{\mu\nu}=(g_{\mu\nu}-k_{\mu}k_{\nu}/k^{2})$, and $R_{\mu}=(u_{\mu}-k_{\mu}(uk)/k^{2})$; where $u_{\mu}$ is the four-velocity of the medium, and $F_{\mu\nu}$ is the electromagnetic tensor. We express tensor quantities in Minkowski metric, in the form $g_{\mu\nu}=(1,1,1,-1)$, understanding the order $\mu,\nu=1,2,3,0$ as it was agreed after Eq. (\ref{ecuacion/SD}). The tensors Eq. (\ref{tens-basicos-S}) are symmetric in the indexes $\mu$, $\nu$ while the following ones are antisymmetric

\begin{eqnarray}
\Psi_{\mu\nu}^{(5)}&=&(uk)[k_{\mu}(Fk)_{\nu}-k_{\nu}(Fk)_{\mu}+k^{2}F_{\mu\nu}]\nonumber\\
\Psi_{\mu\nu}^{(6)}&=&u_{\mu}(Fk)_{\nu}-u_{\nu}(Fk)_{\mu}+(uk)F_{\mu\nu}.
\end{eqnarray}\label{tens-basicos-A}

We introduce a set of orthonormal vectors which are the
  eigenvectors of $\Pi_{\mu\nu}$ in the limit $\mu=0$ and
  $T=0$

\begin{eqnarray}\label{vect-orton}
c_{\mu}^{(1)}&=& R_{1}(F^{2}k)_{\mu}k^{2}-k_{\mu}(kF^{2}k),\quad c_{\mu}^{(2)}=R_{2}(F^{*}k)_{\mu}\nonumber\\
c_{\mu}^{(3)}&=& R_{3}(Fk)_{\mu},\quad c_{\mu}^{(4)}= R_{4}k_{\mu},
\end{eqnarray}

\noindent where $R_{i}, (i=1,2,3,4)$ are normalization parameters, and $F^{*}_{\mu\nu}$ is the dual of the electromagnetic tensor
$F_{\mu\nu}$. Using these vectors we can obtain the scalars

\begin{eqnarray}\label{escalares}
  p&=& c^{(1)\mu}\Pi_{\mu}^{\nu}c^{(1)}_{\nu},\quad s=c^{(2)\mu}\Pi_{\mu}^{\nu}c^{(2)}_{\nu}\\
  t&=& c^{(3)\mu}\Pi_{\mu}^{\nu}c^{(3)}_{\nu},\quad r=c^{(3)\mu}\Pi_{\mu}^{\nu}c^{(1)}_{\nu}
\end{eqnarray}

and the pseudoscalars

\begin{subequations}\label{psudoescalares}
\begin{eqnarray}
  q&=&c^{(2)\mu}\Pi_{\mu}^{\nu}c^{(1)}_{\nu},\quad v=c^{(2)\mu}\Pi_{\mu}^{\nu}c^{(3)}_{\nu}
\end{eqnarray}
\end{subequations}

From the Eq. (\ref{diagonalize form of the polarization tensor}) we can find the polarization properties of three electromagnetic eigenmodes of $\Pi_{\mu\nu}$ \cite{Hugo2,Shabad1}. In the case of propagation along the magnetic field \textbf{B} in a magnetized medium ($k_{\perp}=0$), the eigenmodes of $\Pi_{\mu\nu}$, are

\begin{eqnarray}\label{eigenmodes}
b^{(2)}_{\mu}&=&ac^{(2)}_{\mu}\hspace{1mm}e^{i(k_{3}x_{3}-\omega t)}\nonumber\\
b^{(1,3)}_{\mu}&=&b(c^{(1)}_{\mu}\pm ic^{(3)}_{\mu})\hspace{1mm}e^{i(\textbf{k}_{\bot}\cdot\textbf{r}_{\bot}-\omega t)},
\end{eqnarray}

\noindent with eigenvalues $\eta^{(2)}=s$ and $\eta^{(1,3)}=t\pm \sqrt{-r^2}$ respectively. Here $\textbf{r}_{\bot}$ is the coordinate vector in the plane-$(x,y)$, and $a,b$ are parameters, which have dimensions of vector potential. The electric and magnetic fields associated by these modes are obtained from the equations:

\begin{equation}\label{ecuaciones-campo electrico-magnetico}
  \textbf{E}^{(i)}=-\frac{\partial\textbf{b}^{(i)}}{\partial
x_{0}}-\frac{\partial\ b^{(i)}_{0}}{\partial\textbf{x}},\hspace{3mm} \textbf{H}^{(i)}=\nabla\times \textbf{b}^{(i)},
\end{equation}

\noindent with $(i=1,2,3)$. For the case of
$C$-symmetric $(\mu=0)$, the mode $b_{\mu}^{(3)}$ is a transverse
plane polarized wave, whose electric unit vector is
$\textbf{E}_{u}^{(3)}=\textbf{e}_{\perp}\times \textbf{e}_{3}$ ,
orthogonal to the plane $(\textbf{B}, \textbf{k})$. Where we are defining $\textbf{e}_{\perp}=\frac{\textbf{k}_{\perp}}{k_{\bot}}$ and
$\textbf{e}_{3}=\frac{\textbf{k}_{3}}{k_{3}}$ as the transverse and
parallel unit vectors respectively. The mode $b_{\mu}^{(2)}$ is pure electric and
longitudinal with $\textbf{E}_{u}^{(2)}=\textbf{e}_{B}$ (we recall that $\textbf{e}_{B}=\textbf{B}/B$ is a pseudovector), whereas
$b_{\mu}^{(1)}$ is transverse $E_{u}^{(1)}=\textbf{e}_{\perp}$.  In this $C$-symmetric case  $\eta^{(1)}=\eta^{(3)}$ ,
and the circular polarization unit vectors $(\textbf{E}_{u}^{(1)}\pm
i\textbf{E}_{u}^{(3)})/\sqrt{2}$ are common eigenvectors of $\Pi_{ij}$
and of the rotation generator matrix $A^{3ij}$ .

In the case of $C$-non-symmetric $(\mu\neq 0)$. The second mode
$b_{\mu}^{(2)}$ is the same pure longitudinal wave that in the $C$-symmetric case. The
transverse modes $b_{\mu}^{(1,3)}$ describe circularly polarized
waves in the plane orthogonal to $\textbf{B}$  having different
eigenvalues, typical of Faraday effect.

\subsection{Calculation of $Im[s]$}
\label{sec9}

The denominator $D$ of the integral $s$ (Eq. (\ref{escalar-s}), which have singularities due to $D$) given by:

\begin{equation}\label{Denominador-D}
  D=4z_{1}p_{3}(p_{3}+k_{3})+z^{2}_{1}-4 \omega^{2}\varepsilon^{2}_{n,0},
\end{equation}

\noindent where $z_{1}=k_{3}^{2}-\omega^{2}$ and $\varepsilon_{n,0}^{2}=m^{2}+2enB$, it can be written in the form symmetric under the exchange $\varepsilon_{q}\leftrightarrow \varepsilon_{q^{\prime}}$, $\omega\leftrightarrow-\omega$ \cite{Hugo3}

\begin{multline}\label{denomonador-DI}
 D^{-1}\hspace{-0.1cm}=\hspace{-0.1cm} \frac{1}{8\varepsilon_{q^{\prime}}\varepsilon_{q}\omega} \hspace{-0.1cm}(\frac{1}{\varepsilon_{q^{\prime}}-\varepsilon_{q}-\omega+i\epsilon}
  -\frac{1}{\varepsilon_{q^{\prime}}-\varepsilon_{q}+\omega+i\epsilon} \\
 -\frac{1}{\varepsilon_{q^{\prime}}+\varepsilon_{q}-\omega+i\epsilon}+
\frac{1}{\varepsilon_{q^{\prime}}+\varepsilon_{q}+\omega+i\epsilon}),
\end{multline}

\noindent where $\varepsilon_{q^{\prime}}=\sqrt{(p_{3}+k_{3})^2+m^2+2enB}$ and $\varepsilon_{q}=\sqrt{p_{3}^{2}+m^2+2enB}$, with $q=(n,p_{3})$. The first pair of singularities  are related to excitation of particles to higher energies and the second two are connected to the pair creation. We have added an infinitesimal positive imaginary part $i\epsilon$ to $\omega$, and by using the relation

\begin{equation}\label{delta}
   \frac{1}{s-\omega-i\epsilon}=P\frac{1}{s-\omega}+i\pi\delta(s-\omega),
\end{equation}

\noindent where $P$ corresponds to the principal value in the expression, we get for the imaginary  part of  $D^{-1}$ \cite{Hugo3}

\begin{multline}\label{ImD}
  Im D^{-1}\hspace{-0.1cm}=\hspace{-0.1cm} \pm \frac{\pi}{8\varepsilon_q\varepsilon_{q^{\prime}}\omega}  \hspace{-0.1cm}[\delta(\varepsilon_{q^{\prime}}-\varepsilon_q\mp \omega)+\delta(\varepsilon_{q^{\prime}}-\varepsilon_q\pm \omega) \\
 -\delta(\varepsilon_{q^{\prime}}+\varepsilon_q\mp \omega)],
\end{multline}

\noindent where the $\pm$ signs applies respectively to $\omega\gtrless0$. We can use now Eq. (\ref{ImD}) to obtain the imaginary part the escalar $s$ (Eq. (\ref{escalar-s}))
according to the relation

\begin{equation}
\int_{-\infty}^{\infty}dp_3f(p_3)\delta(g(p_3))=\sum_m\frac{f(p_3^m)}{\mid g^{\prime}(p_3^m)\mid}\label{formula},
\end{equation}

\noindent where $p_3^m$, with $m=(1,2)$ are the roots of $g(p_3)=0$. It may be easily shown that while $p_3$ runs within $(-\infty<p_{3}<\infty)$, the denominator of the expression Eq. (\ref{escalar-s}) may vanish only for real $z_{1}$ \cite{Hugo3}. Thus, the integral in Eq. (\ref{escalar-s}) represents an analytic function in the $z_{1}$ plane except possible singularities located somewhere on the real axis, which corresponds with the absorption region ($Im[\Pi_{33}]$ is responsible of absorption process for the longitudinal mode), where

\begin{equation}\label{momentum}
  p_{3}^{(1,2)}=\frac{-k_{3}z_{1}\pm \omega\Lambda}{2z_{1}},
\end{equation}

\noindent are the roots of denominator in Eq. (\ref{escalar-s}) \cite{Hugo3} and $\Lambda=\sqrt{z_{1}(z_{1}+4\varepsilon^{2}_{n,0})}$. In our case $g(p_3)=\omega\pm(\varepsilon_{q^{\prime}}\pm \varepsilon_{q})$, thus

\begin{equation}\label{balance energia/momentum}
  \omega=\varepsilon_{q^{\prime}}\pm \varepsilon_{q},\hspace{2mm}k_{3}=p^{\prime}_{3}\pm p_{3},
\end{equation}

\noindent and the corresponding values of the energies are given by

\begin{equation}\label{energia-ecxitacion}
  \varepsilon_{r}=\frac{-\omega z_{1}+|k_{3}|\Lambda}{2z_{1}}
\end{equation}
\begin{equation}\label{energia-cracion}
  \varepsilon_{s}=\frac{\omega z_{1}+|k_{3}|\Lambda}{2z_{1}}
\end{equation}

\noindent where $r,s=(n,\omega,k_{3})$. The $\pm$ signs in Eq. (\ref{balance energia/momentum}) corresponds to the pair creation $(\varepsilon_{s})$ and excitation cases $( \varepsilon_{r})$ respectively. By substituting these expressions it is easy to obtain:

\begin{equation}
\mid \frac{d}{dp_3}(g(p_3))\mid=\frac{\Lambda}{2\varepsilon_q^m\varepsilon_{q^{\prime}}^m}
\label{resultado}.
\end{equation}

In the evaluation of the integral Eq. (\ref{escalar-s}) containing the second delta Eq. (\ref{ImD}), the following exchange is made $p_{3}+k_{3}\leftrightarrow -p_{3}$, $n^{\prime}\leftrightarrow n$.

\section*{References}
\label{sec10}

\end{document}